\documentclass[12pt]{iopart}

\usepackage{graphicx}
\usepackage{hyperref}

\usepackage{amssymb}
\def\sgn{\,\mbox{sgn}\,}
\newcommand{\R} {\mbox{Re}\,}
\newcommand{\I} {\mbox{Im}\,}

\def\Xint#1{\mathchoice
   {\XXint\displaystyle\textstyle{#1}}%
   {\XXint\textstyle\scriptstyle{#1}}%
   {\XXint\scriptstyle\scriptscriptstyle{#1}}%
   {\XXint\scriptscriptstyle\scriptscriptstyle{#1}}%
   \!\int}
\def\XXint#1#2#3{{\setbox0=\hbox{$#1{#2#3}{\int}$}
     \vcenter{\hbox{$#2#3$}}\kern-.5\wd0}}

\def\dashint{\Xint-}

\newcommand{\la}{\label}
\newcommand{\be}{\begin{equation}}
\newcommand{\ee}{\end{equation}}
\newcommand{\bea}{\begin{eqnarray}}
\newcommand{\eea}{\end{eqnarray}}

\newcommand{\p}{\partial}
\newcommand{\ba}{\begin{align}}
\newcommand{\ea}{\end{align}}
\newcommand{\1}{\frac{1}{2}}

\def\Xint#1{\mathchoice
   {\XXint\displaystyle\textstyle{#1}}%
   {\XXint\textstyle\scriptstyle{#1}}%
   {\XXint\scriptstyle\scriptscriptstyle{#1}}%
   {\XXint\scriptscriptstyle\scriptscriptstyle{#1}}%
   \!\int}
\def\XXint#1#2#3{{\setbox0=\hbox{$#1{#2#3}{\int}$}
     \vcenter{\hbox{$#2#3$}}\kern-.5\wd0}}

\def\dashint{\Xint-}

\begin{document}


\title[Integrable hydrodynamics of Calogero-Sutherland \ldots]{Integrable hydrodynamics of Calogero-Sutherland model: Bidirectional Benjamin-Ono equation.
 }

\author{Alexander G. Abanov
}
\address{Department of Physics and Astronomy,
Stony Brook University,  Stony Brook, NY 11794-3800.}

\author{Eldad Bettelheim}
\address{Racah Institute of Physics, The Hebrew University of
Jerusalem, Safra Campus, Givat Ram, Jerusalem, Israel 91904.}

\author{Paul Wiegmann}
\address{James Franck Institute
of the University of Chicago,
5640 S.Ellis Avenue, Chicago, IL 60637.
}

\begin{abstract}

We develop a hydrodynamic description of the classical
Calogero-Sutherland liquid: a Calogero-Sutherland model with an
infinite number of particles and a non-vanishing density of
particles. The hydrodynamic equations, being written for the
density and velocity fields  of  the liquid, are shown to be a
bidirectional analogue of Benjamin-Ono equation. The latter is known
to describe internal waves of deep stratified fluids.  
We show that the bidirectional Benjamin-Ono equation appears as a real
reduction of the modified KP hierarchy.
We derive the Chiral Non-linear Equation which appears
as a chiral reduction of the bidirectional equation. The
conventional Benjamin-Ono equation is a degeneration of the Chiral
Non-Linear Equation at large density.  We construct multi-phase
solutions of the bidirectional Benjamin-Ono equations and of the
Chiral Non-Linear equations.
\end{abstract}

\maketitle


\tableofcontents





\section{Introduction}

The Calogero-Sutherland model (CSM)
\cite{Calogero-1969,Sutherland-1971} describes particles moving on a
circle and interacting through an inverse $\sin$-square potential.
The  Hamiltonian of the model reads \be
 \label{CSM}
    {\cal H}_{CSM} = \frac{1}{2}\sum_{j=1}^{N}p_j^{2}
    +\1\left(\frac{\pi}{L}\right)^{2}\sum_{j,k=1; j\neq k}^{N}
    \frac{g^{2}}{\sin^{2}\frac{\pi}{L}(x_{j}-x_{k})},
\ee where $x_{j}$ are coordinates of $N$ particles, $p_{j}$ are
their momenta, and $g$ is the coupling constant. We took the mass of
the particles to be unity. The momenta $p_{j}$ and coordinates
$x_{j}$ are canonically conjugate variables.

The model (classical and quantum) occupies an exceptional place in
physics and mathematics and has been studied extensively. It  is
completely integrable.
Its solutions can be written down explicitly as finite dimensional
determinants (for review see
\cite{1981-OlshanetskyPerelomov-classical}).

In the limit of a large period  $L\to\infty$ the CSM degenerates to
its rational version -- Calogero (aka Calogero-Moser) model (CM)
where the pair-particle interaction is $1/{x^{2}}$.
\footnote{In the rational case one usually adds a harmonic
potential, $\1\omega^2 \sum_i x_i^2$,   to the Hamiltonian to
prevent particles from escaping. This addition  does not destroy the
integrability of the system \cite{Sutherland-1971}.} The CSM itself
is a degeneration of the elliptic Calogero model, where the pair
particle interaction is given by the Weierstrass $\wp $-function of
the distance. In this paper we discuss the classical  trigonometric
model (\ref{CSM}) commenting on the rational limit  when
appropriate.

We are interested in describing a Calogero-Sutherland {\it liquid},
i.e., the system (\ref{CSM}) in thermodynamic limit when $N\to
\infty$ and $L\to \infty$ while the average density $N/L$ is kept
constant.  We assume that the limit exists and that in this limit  a
microscopic density and current fields 
\bea
    \rho(x,t) &=& \sum_{j=1}^{N}\delta(x-x_{j}(t)), 
 \label{3} \\
    j(x,t) &=& \sum_{j=1}^{N} p_{j}(t) \delta(x-x_{j}(t))
\eea 
are smooth single-valued  real periodic functions with a period $L$
equal to the period of  the potential  \footnote{It is likely that
there are classes of solutions of the CSM, whose thermodynamic limit
consists of a number of  interacting liquids. In this case the
microscopic density  give rises to a number of functions in the
continuum - the densities of the distinct interacting liquids. In
this paper we consider a class of solutions which leads to a single
liquid.}. In this case the system will be described by hydrodynamic
equations written on the density field $\rho(x,t)$ and the velocity
field $v(x,t)$. The velocity is defined as $j=\rho v$.

The hydrodynamic approach is a  powerful tool to study  the
evolution of smooth features with typical size much larger than the
inter-particle distance. Apart from application to the CSM, the
hydrodynamic equations obtained in this paper are interesting
integrable equations. We show that they are new  real reductions of
the modified Kadomtzev-Petviashvili  equation (MKP1).

In this paper we consider  a classical system, however the approach
developed below can be extended to the quantum  case
$\{p_{j},\,x_{k}\}=\delta_{jk}\rightarrow [p_{j},\,x_{k}]=i\hbar
\delta_{jk}$ almost without changes. For a brief description of the
hydrodynamics of the quantum system see
Ref.~\cite{2005-AbanovWiegmann}.  The hydrodynamics of the quantum
Calogero model has been studied previously
\cite{AJL-1983,AndricBardek-1988} in the framework of the {\it
collective field theory} and some of the results below can be obtained in a classical
limit (see \cite{1995-Polychronakos}) of  the quantum counterparts of Refs.
\cite{AJL-1983,AndricBardek-1988}.

The outline of this paper is the following. In
Sec.~\ref{particlespoles} we parameterize the particles of CSM
as poles of auxiliary complex fields so that the motion of particles
is encoded by evolution equations for fields. In
Sec.~\ref{hydrodynamiclimit} we derive a hydrodynamic limit of these
equations - continuity and Euler equations with a particular form of
specific enthalpy. We will refer to these equations as to the
bidirectional Benjamin-Ono equation or 2BO. We present the
Hamiltonian form of 2BO in Sec.~\ref{sec:HformdBO}. In
Sec.~\ref{sec:BilformdBO} we discuss the bilinear form  of 2BO and
its relation to MKP1.  In Sec.~\ref{sec:chiral} we obtain the Chiral
Non-Linear equation (CNL)  - chiral reduction of 2BO and discuss some of
its properties. In Sec.~\ref{sec:Multi-phase} we construct
multi-phase and multi-soliton solutions of 2BO and CNL as a real reduction
of MKP1. These solutions correspond to collective excitations of the
original many-body system.  Some technical points are relegated to
the appendices.

\section{Particles as poles of meromorphic functions}
 \la{particlespoles}

The Equations of motion of the CSM  are readily obtained from the
Hamiltonian (\ref{CSM}) 
\bea
    \dot{x}_{j} &=& p_{j},
 \la{csmeq1} \\
    \dot{p}_{j} &=& -g^{2} \frac{\partial}{\partial x_{j}}
    \sum_{k=1\, (k\neq j)}^{N}\left(\frac{\pi}{L}\cot\frac{\pi}{L}(x_{j}-x_{k})\right)^{2}.
 \la{csmeq2}
\eea
We rewrite this system in an equivalent way as
\bea
    i\frac{\dot{w}_{j}}{w_{j}} &=& \frac{g}{2}\left(\frac{2\pi}{L}\right)^{2}
    \left(\sum_{k=1}^{N}\frac{w_{j}+u_{k}}{w_{j}-u_{k}}
    - \sum_{k=1\,(k\neq j)}^{N} \frac{w_{j}+w_{k}}{w_{j}-w_{k}}\right),\quad j=1,\dots, N
 \la{pmotx}
 \\
    -i\frac{\dot{u}_{j}}{u_{j}} &=&  \frac{g}{2}\left(\frac{2\pi}{L}\right)^{2}
    \left(\sum_{k=1}^{N}    \frac{u_{j}+w_{k}}{u_{j}-w_{k}}
    - \sum_{k=1\,(k\neq j)}^{N}\frac{u_{j}+u_{k}}{u_{j}-u_{k}}\right),\quad j=1,\dots, N,
 \la{pmoty}
\eea 
where $w_{j}(t) = e^{i\frac{2\pi}{L}x_{j}(t)}$   are complex
coordinates  lying  on a unit circle, while    $u_{j}(t) =
e^{i\frac{2\pi}{L}y_{j}(t)}$ are auxiliary coordinates.  Indeed,
differentiating (\ref{pmotx}) with respect to time and using
(\ref{pmotx},\ref{pmoty}) to remove first derivatives in time one
obtains equations equivalent to (\ref{csmeq1},\ref{csmeq2}).

We note that while the coordinates $x_j$ are real, i.e., $|w_j|=1$,
the auxiliary coordinates, $y_{j}(t)$, are necessarily complex.
Given initial data as real positions and velocities $x_{j}(0)$ and
$\dot{x}_{j}(0)$ one can find complex $y_{j}$ from (\ref{pmotx}) and
then initial complex velocities $\dot{y}_{j}(0)$ from (\ref{pmoty}).
Once  $x_{j}$ and $\dot{x}_{j}$ are  chosen to be real they will
stay real at later times, even though  coordinates $y_i$ are moving in a complex plane.

The coordinates $w_{j}(t)$ and $u_{j}(t)$ determine an evolution of
two functions 
\bea
    u_1(w) &=& g\frac{\pi}{L}\sum_{j=1}^{N}\frac{w+w_{j}}{w-w_{j}}
    =-i g \sum_{j=1}^{N} \frac{\pi}{L} \cot\frac{\pi}{L}(x-x_{j}),\quad w= e^{i\frac{2\pi}{L}x},
 \la{u-pa} \\
    u_0(w) &=& -g\frac{\pi}{L}\sum_{j=1}^{N}\frac{w+u_{j}}{w-u_{j}}
    =i g \sum_{j=1}^{N} \frac{\pi}{L} \cot\frac{\pi}{L}(x-y_{j}), \quad w= e^{i\frac{2\pi}{L}x}.
 \la{u+pa}
\eea 
The latter functions play a major role in our approach. These are rational
functions of $w$ regular at infinity and having particle coordinates
as simple poles with equal residues $2\pi g/L$.

The  condition that the coordinates of particles $x_{j}$ are real
yields Schwarz reflection condition  for the function $u_1$ with
respect to the unit circle \bea
    \overline{u_1(w)} = -u_1(1/\bar{w})
    \qquad \mbox{or}
     \qquad
    \overline{u_1(x)} = - u_1(\bar{x}),
 \la{Schwarz}
\eea where bar denotes complex conjugation. The values of $u_1(w)$
in the interior and exterior of a unit circle are related by Schwarz
reflection.

Comparing (\ref{pmotx}), (\ref{csmeq1}) and (\ref{u+pa}) we notice
that while the function $u_{1}(w)$ encodes the positions of particles
$w_{j}$, the  function $u_0(w)$ encodes the momenta of particles as
its values at particle positions $w_{j}$ \be
    p_{j} = u_0(w_{j}) + g\frac{\pi}{L} \sum_{k=1\,(k\neq j)}^{N} \frac{w_{j}+w_{k}}{w_{j}-w_{k}}.
 \la{csmmomenta}
\ee We notice here that the positions of the particles fully
determine the imaginary part of the field $u_0$ on a unit circle.
Indeed, we have from (\ref{csmmomenta}) \be
    \I u_0(x_{j}) = g {\sum_{k\neq j}}\frac{\pi}{L}\cot \frac{\pi}{L}(x_{j}-x_{k}).
 \la{u+real1}
\ee

We now introduce complex functions \be
    u=u_0+u_1,\quad \tilde u=u_0-u_1.
 \la{decomp}
\ee
One can show that they obey the  equation
\be
 \la{2BO}
    u_{t}+\partial_{x}\left[\frac{1}{2}u^{2}+i\frac{g}{2} \partial_{x}\tilde u \right] =0.
\ee Indeed substituting the  \textit{pole ansatz}
(\ref{u-pa},\ref{u+pa}) into (\ref{2BO}) and comparing the residues
at  poles $w_{j}$ and $u_{j}$ one arrives at
(\ref{pmotx},\ref{pmoty}).

The equation (\ref{2BO}) connects two complex functions $u_0$ and
$u_1$.  The equation is equivalent  to the {\it modified
Kadomtzev-Petvisashvili} equation (or simply MKP1). We will discuss
its relation to MKP1  in Sec.~\ref{sec:BilformdBO}.

However,  being complemented by the Schwarz reflection condition
(\ref{Schwarz}), analyticity requirements, and an additional reality requirement
it becomes an equation uniquely determining $u_0$ and $u_1$ through their initial data.

The analyticity requirements read: $u_0(w)$ is analytic in a
neighborhood of a unit circle $|w|=1$, while $u_1$ is analytic
inside $|w|<1$ and outside $|w|>1$ of the unit circle,  approaching
a constant at $w\to\infty$. An additional reality requirement is the
relation between the imaginary part of $u_{0}$ on a unit circle and
$u_{1}$ stemming from the condition (\ref{u+real1}). We formulate
and discuss these conditions in Sec.~\ref{subsec:2BO} and
Sec.~\ref{sec:BilformdBO}.

We will refer to the equation (\ref{2BO}) as the bidirectional
Benjamin-Ono equation (2BO). It is a bidirectional (having both
right and left moving waves) generalization of the conventional {\it
Benjamin-Ono} equation (BO) arising in the hydrodynamics of
stratified fluids \cite{AblowitzClarkson-book}. We discuss its
hydrodynamic form in the next section.

The solution of (\ref{2BO}) given by (\ref{u-pa},\ref{u+pa}) is the CSM many body system with a
finite number of particles (\ref{CSM}). Other solutions describe CSM
fluids. They are the central issue of this paper.

To conclude this section we make the following comment. The function
$u_1$ can be expressed solely in terms of the microscopic density of
particles (\ref{3}) as \be
    u_1(w) = -\pi g \oint \frac{d\zeta}{2\pi i \zeta}\, \frac{\zeta+w}{\zeta-w}\,\rho(\zeta).
 \la{u-def}
\ee The integral in this formula goes  over the unit circle
$\zeta=\exp\left(i\frac{2\pi}{L}x\right)$. In the following we will
denote for brevity $\rho(\zeta)$ as $\rho(x)$, when $\zeta$ lies on
a unit circle $\zeta = e^{i\frac{2\pi}{L}x}$.  The density itself
can be obtained as a difference of  limiting values of the field
$u_1$ at the real $x$ (on the unit circle). The discontinuity of
$u_{1}$ on the unit circle gives a microscopic density (\ref{3})  of particles
\bea
    u_1(x+ i0)-u_1(x- i0) &=& -2\pi g\rho(x), \quad \I x=0,\quad 0<\R x<L.
 \label{discont}
\eea

\section{Hydrodynamics of Calogero-Sutherland liquid}
 \la{hydrodynamiclimit}

\subsection{Density and velocity}

We assume that in the thermodynamic limit $N, L\to\infty$, $N/L=const$  the poles
of the function $u_1$ are distributed along the real axis with a
smooth density  $\rho(x)$ and consider a  complex field $u_1(w)$
given by formula (\ref{u-def}). Notice that $u_1(w)$ defined by
(\ref{u-def}) is analytic everywhere outside of the real axis of $x$
(everywhere off the unit circle in $z$-plane) approaching a constant
as $z\to\infty$. It also satisfies the reality condition
(\ref{Schwarz})  (the density $\rho(x)$ is real). In the thermodynamic limit the function $u_1$
is not a rational  function anymore. It is discontinuous across
the real axis with the discontinuity related to the density of
particles by (\ref{discont}). The value of the field $u_1(x)$ on a
real axis (on a unit circle in $z$ plane)  depends on whether one
approaches the real axis from above or below (unit circle from the
interior $z\to e^{i\frac{2\pi}{L}(x+i0)}$ or from the exterior $z\to
e^{i\frac{2\pi}{L}(x-i0)}$). More explicitly, we have from
(\ref{u-def}) \be
    u_1(x\pm i0)
    =  \pi g (\mp\rho + i\rho^{H}).
 \la{u-hydr}
\ee
The superscript $H$ in the second term of (\ref{epsrho0}) denotes the Hilbert transform and is defined as (see \ref{app-hilbert} for definitions and some properties of the Hilbert transform)
\be
    f^{H}(x) =\dashint_{0}^{L}\frac{dy}{L}\, f(y)\cot\frac{\pi}{L}(y-x).
 \la{HtransC}
\ee

We also assume that in $N\to\infty$ limit the complex field $u_0(w)$
remains  analytic in the vicinity of the real axis in $x$-plane
(i.e., in the vicinity of a unit circle in $z$-plane).

The 2BO (\ref{2BO}) does not explicitly depend on the number of
particles $N$. It holds also in thermodynamic limit $N, L\to\infty$,
$N/L=const$, however solutions describing a liquid are not rational
functions any longer.

We can use 2BO to define velocity through the \textit{continuity
equation} \bea
    \rho_{t} &+&\partial_{x}(\rho v) = 0.
 \la{continuity}
\eea

The discontinuity of the complex field $u(x)$ (\ref{decomp}) across
the real axis, as well as a discontinuity of the field $u_1$ (see
(\ref{discont})) is the density \be
    u(x+i0)-u(x-i0) = u_1(x+i0)-u_1(x-i0) = -2\pi g \rho(x).
 \la{udiscont}
\ee
Differentiating (\ref{udiscont}) with respect to time and using
2BO (\ref{2BO}) we obtain the continuity equation and identify the
\textit{velocity field} $v(x)$ as \bea
    v(x) &=& u_0(x) +\frac{1}{2}\left(u_1(x+i0)+u_1(x-i0)\right)
    -ig\partial_{x}\log\sqrt\rho(x)
 \nonumber \\
    &=& u_0(x) +ig\left(\pi \rho^{H}(x) -\partial_{x}\log\sqrt\rho(x)\right)
\eea
or
\be
    u_0(x) = v -ig\left(\pi \rho^{H} - \partial_{x}\log\sqrt\rho\right).
 \la{u+hydr}
\ee
Since $v(x)$ is a real field (\ref{u+hydr}) provides  a reality
condition analogous to (\ref{u+real1}). Indeed, one can see from
(\ref{u+hydr}) that \be
    \I u_0(x) = -g\left(\pi  \rho^{H}-\partial_{x}\log\sqrt\rho\right),
 \la{u+real2}
\ee i.e., the imaginary part of $u_0(x)$ is completely determined by
the density of particles or equivalently by the field $u_1$. It is
also convenient to have an expression for $u(x)$ on a real axis \bea
    u(x\pm i0) = v + g\left(\mp \pi \rho +i\partial_{x}\log\sqrt\rho\right).
 \la{ux}
\eea
It has the same discontinuity across the real axis as  $u_1(x)$.

\subsection{Hydrodynamic form of 2BO.}

Now we are ready to cast the equation (\ref{2BO}) into hydrodynamic form.

Taking the real part of 2BO (\ref{2BO}) on the real axis and using
identifications (\ref{u-hydr},\ref{u+hydr}) and the continuity
equation, (\ref{continuity}), after some algebra we arrive at the
\textit{Euler equation} \bea
    v_{t} &+& \partial_{x}\left(\frac{v^{2}}{2} +w(\rho)\right) = 0,
 \la{Euler}
\eea
with specific (per particle) enthalpy or chemical potential\footnote{The specific enthalpy and chemical potential are identical at zero temperature.} given by
\be
	w(\rho) =\frac{1}{2}(\pi g \rho)^{2}
	-\frac{g^2}{2}\frac{1}{\sqrt{\rho}}\,\partial_x^2\sqrt{\rho}    +\pi g^{2}\rho_{x}^{H}.
 \la{enthalpy}
\ee 
Equations (\ref{continuity},\ref{Euler}) are  the continuity and
Euler\footnote{The eq.~(\ref{Euler}) has a form of an Euler equation for an isentropic flow. Because of the long range character of interactions the enthalpy cannot be replaced by the conventional pressure term $\partial_{x}w(\rho) \to \rho^{-1}\partial_{x}(p(\rho))$ - the standard form of the Euler equation.}  equations of classical Calogero-Sutherland model.  They are
the classical analogues  of quantum hydrodynamic equations that have
been obtained for the quantum CSM in Refs.
\cite{AJL-1983,AndricBardek-1988,Awata} first using collective field
theory approach \cite{JevickiSakita,Sakita-book,Jevicki-1992} and
later by the {\it pole ansatz} similar to the one used above
\cite{2005-AbanovWiegmann}.
It was noticed in \cite{Jevicki-1992} and then in
\cite{1995-Polychronakos} that the system
(\ref{continuity},\ref{Euler},\ref{enthalpy}) has a lot of
similarities with classical Benjamin-Ono equation
\cite{Benjamin-Ono}. The similarities and differences with
Benjamin-Ono equation are discussed below. We will refer to
(\ref{continuity},\ref{Euler},\ref{enthalpy}) as to a hydrodynamic
form of the \textit{bidirectional Benjamin-Ono equation} (2BO).

\subsection
{Bidirectional Benjamin-Ono equation (2BO).}
\la{subsec:2BO}

Let us now summarize the 2BO equation:
 \bea
 \la{2BO1}
    && u_{t}+\partial_{x}\left[\frac{1}{2}u^{2}+i\frac{g}{2} \partial_{x}\tilde u \right] =0,\\
    && u=u_0+u_1,\quad \tilde u=u_0-u_1.
\eea
The functions $u_0$ and $u_1$ are subject to analyticity  conditions
\bea
 \label{A1}
    && u_1(x) \qquad \mbox{- analytic for}\;\; \I(x) \neq 0,
 \\
  \label{A2}
    && u_0(x) \qquad \mbox{- analytic for}\;\; |\I(x)|
    <\epsilon \;\; \mbox{for some}\;\;  \epsilon>0,
\eea
and to reality conditions
\be
 \label{A3}
    \overline{u_1(x)}=-u_1(\bar{x}).
\ee 
In addition, the fact that the equation (\ref{2BO1}) holds in
the upper half plane and in the lower half plane (inside and outside
of the unit circle) yields the condition
 \be    \I[u(x\pm i0)] = \frac{g}{2} \partial_{x} \log \R [u_1(x\pm i0)].
 \la{u+Im}
\ee It also follows from  (\ref{u-hydr},\ref{u+real2},\ref{ux}). The
condition (\ref{u+Im}) looks more ``natural'' in the bilinear
formulation (see eq. (\ref{blreal}) below).

These reality and analyticity conditions reduce two complex fields
$u_0$ and $u_1$ to two real fields - density $\rho(x)$ and velocity
$v(x)$ as (\ref{u-hydr},\ref{u+hydr}). Then, a complex  equation
(\ref{2BO}) defined in both half planes immediately yields the
hydrodynamic equations
(\ref{continuity},\ref{Euler},\ref{enthalpy}). Inversely, knowing
real periodic fields $\rho(x)$ and $v(x)$ one can find fields $u_0,
u_1$ everywhere in a complex $x$-plane.

\subsubsection*{Mode expansion}

The analyticity and reality conditions can be recast in the language
of mode expansions. It follows from (\ref{u-def}) that \be\label{22}
    u_1(w) = \left\{
\begin{array}{lr}
    - \pi g\left(\rho_{0}+2\sum_{n=1}^{\infty}\rho_{ n}w^n\right),\quad |w|<1
 \\
    \pi g\left(\rho_{0}+2\sum_{n=1}^{\infty}\rho_{n}^\dag w^{- n}\right),\quad |w|>1
\end{array}
\right.\ee
where $\rho_{n} =\rho^\dag_{-n}=\int_{0}^{L}\frac{dx}{L}\, \rho(x) e^{-i\frac{2\pi n}{L}x}$ are Fourier components of the density.

The values of the field $u_{1}(w)$ in the upper and lower half-planes are then automatically related by Schwarz reflection (\ref{Schwarz}).

Conversely, the field  $u_0(x)$  being analytic in a  strip around the unit circle is represented by Laurent series
\be
 \la{38}
    u_0(w) =V_0+ \sum_{n=1}^{\infty}\left(a_{n} w^{n}+b_nw^{-n}\right), \quad  |\I\log w|<2\pi \epsilon/L.
\ee

The 2BO equation remains intact in the case of rational
degeneration. Rational degeneration of formulas of the
Sec.~\ref{particlespoles} are obtained by a direct expansion in
$1/L$ . In this limit fields are defined microscopically as $u_1(x)=
-ig\sum_j\frac{1}{x-x_j}$ and $u_0(x)= ig\sum_j\frac{1}{x-y_j}$.

\section{Hamiltonian form of 2BO}
\la{sec:HformdBO}

The 2BO is a Hamiltonian equation. Let us start with its Hamiltonian
formulation in the  hydrodynamic form
$\rho_{t}=\left\{H,\rho\right\},\;v_{t}=\left\{H,v\right\}$ with the
canonical Poisson bracket of density and velocity fields \be
    \left\{\rho(x),v(y)\right\} = \delta'(x-y).
 \la{rhovcom}
\ee
Equations (\ref{continuity},\ref{Euler},\ref{enthalpy}) follow from
\bea
    H&=& \int dx\, \left(\frac{\rho v^{2}}{2}
    +\rho\epsilon(\rho)\right),
 \la{csmrv0} \\
    \epsilon(\rho) &=&\frac{g^2}{2}(\pi \rho^{H} -\partial_{x}\log\sqrt\rho)^{2}.
 \la{epsrho0}
\eea 
Here the ``internal energy'' (\ref{epsrho0}) and the enthalpy
(\ref{enthalpy}) are related by a general formula
$w(\rho)=\frac{\delta }{\delta \rho(x)}\int dx\, \rho
\epsilon(\rho)$.

For references we will give alternative expressions for the
Hamiltonian. Let $\Psi=\sqrt\rho e^{i\vartheta}$ where
$v=g\p_x\vartheta$ then \be
    H= \frac{g^2}{2}\int \left|\partial_{x}\Psi-\pi\rho^{H}\Psi\right|^2 dx,
 \la{hamPsi}
\ee
where $\rho = |\Psi|^{2}$. The Poisson's brackets  for $\Psi(x)$ are canonical:
$\left\{\Psi(x),\Psi(y)\right\}=0$, and
$\left\{\Psi(x),\Psi^{\star}(y)\right\}=\frac{i}{g}\delta(x-y)$. The
equations of motion for $\Psi$ and $\Psi^{\star}$ are 
\be
    \frac{i}{g}\partial_{t}\Psi
    = \left[-\frac{1}{2}\partial_{x}^{2}+\frac{\pi^{2}}{2}|\Psi|^{4}+\pi \left(|\Psi|^{2}\right)^{H}_{x}\right] \Psi
\ee
and its complex conjugate. A simple change of a dependent variable $\Phi=\Psi e^{i\pi\int^{x}dx'\,|\Psi(x')|^{2}}$ leads to
\be
    \frac{i}{g}\partial_{t}\Phi = \left[-\frac{1}{2}\partial_{x}^{2}+i 2\pi \left(|\Phi|^{2}\right)^{+}_{x}\right] \Phi,
 \la{INLS}
\ee 
where $f^{+}$ denotes the function analytical in the upper
half-plane of $x$ defined as $f^{+}=\frac{f-if^{H}}{2}$. One can
recognize in (\ref{INLS}) the intermediate nonlinear Schr\"odinger
equation (INLS) which appeared in Ref.\cite{1995-Pelinovsky} as an
evolution of the modulated internal wave in a deep stratified fluid.
Therefore, one can alternatively think of 2BO as the hydrodynamic
form of (\ref{INLS}) identifying hydrodynamic fields $\rho$ and $v$
to be 
\be
    \Phi = \sqrt{\rho} \exp\left\{\frac{i}{g}\int^{x}dx'\, (v+\pi g\rho)\right\}
 \la{back1}
\ee
or with the field $u(x)$ from (\ref{ux}) as
\be
    ig\partial_{x}\log\Phi^{\star} = u(x-i0).
 \la{back2}
\ee
The Hamiltonian (\ref{csmrv0}) or (\ref{hamPsi}) can be rewritten in terms of $\Phi$ as
\be
    H= \frac{g^2}{2}\int \left|\partial_{x}\Phi-i 2\pi\rho^{+}\Phi\right|^2 dx,
 \la{hamPhi}
\ee 
where $\rho=|\Phi|^{2}$. 
However, the Poisson's brackets for $\Phi$ are not canonical
anymore.\footnote{Simple calculation using (\ref{rhovcom}) gives
$\left\{\Phi(x),\Phi(y)\right\}=\frac{\pi}{g}\Phi(x)\Phi(y)\sgn(x-y)$,
$\left\{\Phi(x),\Phi^{\star}(y)\right\}=\frac{i}{g}\delta(x-y)-\frac{\pi}{g}\Phi(x)\Phi^{\star}(y)\sgn(x-y)$
and similar expressions for complex conjugated fields. One should
think of $\Psi(x)$ as a canonical bosonic field while of $\Phi(x)$ as a classical analogue of a field with fractional statistics.}

2BO is an integrable system.  It has infinitely many integrals of motion. The first three of them follow from global symmetries. They are conventional the number of
particles $N= \int dx\, \rho$, the total momentum $P= \int dx\,
\rho v$, and the total energy $H=\int dx\, \left(\frac{\rho v^{2}}{2}+\rho\epsilon(\rho)\right)$. 
They are conveniently written in terms of the fields $u$ and $\tilde u$ as  
\bea
    I_{1} &=& N = \frac{1}{2\pi g}\oint_{C}dx\, u,
 \la{numdbo} \\
    I_{2} &=& P = \frac{1}{2\pi g}\oint_{C}dx\, \frac{1}{2}u^{2},
 \la{momdbo} \\
    I_{3} &=& 2H = \frac{1}{2\pi g}\oint_{C}dx\,
    \left[\frac{1}{3}u^{3} +i\frac{g}{2} u \partial_{x}\tilde  u\right],
 \la{hamdbo}
\eea
where the integral is taken over the both sides of the unit circle.
(``double'' contour $C$ shown in Fig. \ref{fig:leftright}).
For more details on conserved integrals see \ref{app:int}.
\begin{figure}
\bigskip
\begin{center}
 \includegraphics[width=6cm]{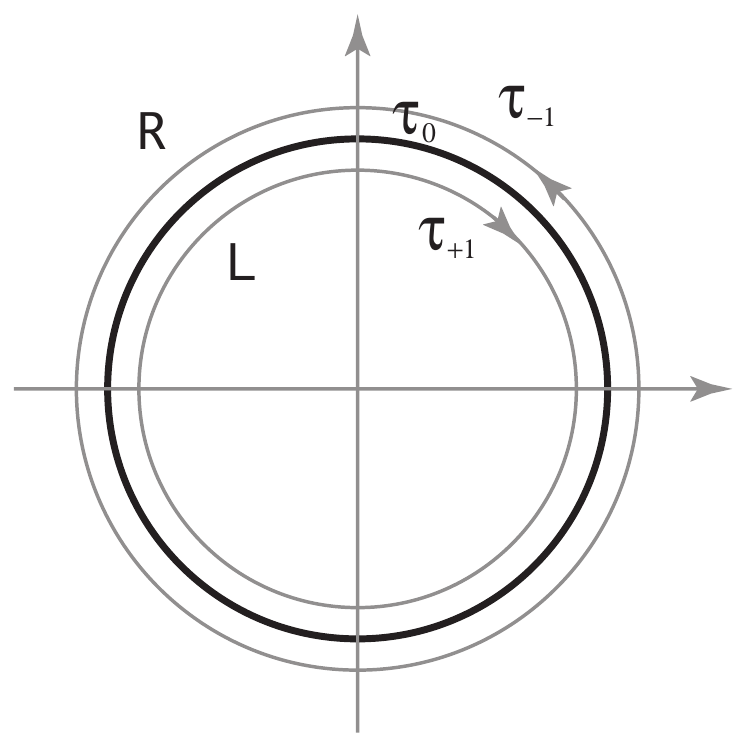}
\end{center}
\caption{Contour $C$ surrounding the unit circle is shown together with our conventions in defining right and left fields.}
\la{fig:leftright}
\end{figure}

The Poisson bracket for the fields $u_0(w)$ and $u_1(w)$ can be
easily obtained from from  (\ref{u-hydr},\ref{u+hydr}) and
(\ref{rhovcom}) by analytic continuation. We find that
$\left\{u_0(w),u_0(w')\right\} =\left\{u_1(w),u_1(w')\right\}=0$ and
\bea
    \left\{u_0(w),u_1(w')\right\} &=&
    ig\left(\frac{2\pi}{L}\right)^{2} \frac{ww'}{(w-w')^{2}}
    = i g \partial_{x} \frac{\pi}{L}\cot \frac{\pi}{L}(x-y).
 \la{PB01}
\eea

\section{Bilinearization and relation to MKP1 equation}
\la{sec:BilformdBO}

The equations described in the previous section, their integrable
structures and their connection to integrable hierarchies are the
most transparent in  bilinear form.

Let us introduce tau-functions $\tau_{0}$ and $\tau_{1}$ as
\bea
    u_0 &=& ig \partial_{x}\log\tau_{0},
 \la{tau01} \\
    u_{1} &=& -ig \partial_{x}\log\tau_{1}.
 \nonumber
\eea
It can be easily checked that the 2BO (\ref{2BO}) can be rewritten as an elegant bilinear Hirota equation on  $\tau$-functions:
\be
    \left(iD_{t}+\frac{g}{2}\, D_{x}^{2}\right) \tau_1 \cdot \tau_0=0.
 \la{2BO-hirota}
\ee
Here we used the Hirota derivative symbols defined as
\bea
    D_{x}^{n}f(x)\cdot g(x) \equiv \lim_{y\to x}(\partial_{x}-\partial_{y})^{n}f(x)g(y).
\eea
For example,
\bea
    D_{t}f \cdot g &=& (\partial_{t}f) g-f(\partial_{t}g),
 \nonumber \\
    D_{x}^{2} f \cdot g  &=& (\partial_{x}^{2}f) g-2(\partial_{x}f)(\partial_{x}g)
    +f(\partial_{x}^{2}g),
\eea
etc.

We emphasize that the bilinear equation holds on both sides of the unit circle. Introducing notations
\be
	\tau_{\pm 1}=\tau_1(x\pm i0)
\ee
we can rewrite the equation as
\bea
    &&\left(iD_{t}+\frac{g}{2}\, D_{x}^{2}\right) \tau_{+1} \cdot \tau_0=0, \\
    &&\left(iD_{t}+\frac{g}{2}\, D_{x}^{2}\right) \tau_{-1} \cdot \tau_{0=}\left(-iD_{t}+\frac{g}{2}\, D_{x}^{2}\right) \tau_{0} \cdot \tau_{-1}=0.
 \la{2BO-hirota1}
\eea

Equation (\ref{2BO-hirota}) is the modified Kadomtsev-Petviashvili
equation (MKP1). MKP1 contains two independent functions $\tau_1$
and $\tau_0$ and is formally not closed.  The analyticity and
reality conditions (\ref{A1}-\ref{u+Im})   stemming from the fact
that all solutions are determined by two real functions $\rho(x,t)$
and $v(x,t)$, close the equation. Under these conditions the
equations can be seen as a real reduction of MKP1. Let us formulate
these conditions in terms of tau-functions.

The first requirement is that $\tau_{\pm 1}$ is analytic and does
not have zeros for $\I x>0$ ($<0$) after analytic continuation. Also
$\tau_0$ should be analytic and should not have zeros in the
vicinity of the real axis, i.e., for $|\I x|<\epsilon$ for some
$\epsilon>0$.

The second requirement is that $\tau_{\pm 1}$ should be related by
Schwarz reflection (\ref{Schwarz}). In terms of tau-functions it
becomes on the unit circle (for real $x$) \be
    \tau_{-1} =\overline{ \tau_{+1}} e^{i\Theta(t)},
 \la{SchwarzTau}
\ee
where a phase $\Theta(t)$  can be any time-dependent function.

The third requirement is related to the fact that $\I u_0$ is a
function of density only and, therefore, can be expressed in terms
of $u_1$ as can be easily seen from (\ref{u-hydr},\ref{u+hydr}).
This condition (\ref{u+Im}) can be written in a bilinear form as
follows \bea
    i D_{x}\tau_{+1}\cdot {\tau_{- 1} }= 2\pi\, \tau_0  \overline{\tau_0}.
\la{blreal} \eea The multiplicative constant in the r.h.s of
(\ref{blreal}) fixes the relative normalization of $\tau_{0}$ and
$\tau_{1}$ and is arbitrary. We have chosen it to be $2\pi$.

Finally, we note  that the pole ansatz solution (\ref{u-pa},\ref{u+pa})
corresponds to the polynomial form of tau-functions with zeros at $w_j$
and $u_j$ 
\bea
    \tau_1(w,t) &=& w^{-N/2} \prod_{j=1}^{N}(w-w_{j}(t)),
 \\
    \tau_0(w,t) &=& w^{-N/2} \prod_{j=1}^{N}(w-u_{j}(t)).
\eea

\section{Chiral Fields and Chiral Reduction}
\la{sec:chiral}

\subsection{Chiral fields and currents}

The 2BO equation can be conveniently expressed through yet another
right and left handed  chiral fields 
\be
 \label{RL}
    J_{R,L} = v \pm g\left[\pi  \rho+\partial_{x}(\log\sqrt\rho)^{H}\right].
\ee 
These fields are real. \footnote{$J_{R,L}$ can be expressed solely in terms of $u_{0}$ field. It is easy to check that (\ref{RL}) is equivalent to $J_{R,L}= \R\left(u_{0}\mp i u_{0}^{H}\right)$.} In terms of them, the 2BO equation (\ref{2BO}) reads
\bea
    \partial_{t}J_{R,L} &+& 
    \partial_{x}\left(\frac{J_{R,L}^{2}}{2}
    \pm \frac{g}{2}\partial_x J_{R,L}^{H}\right) 
 \nonumber \\
    &\mp& 
    g\partial_{x}\left[J_{R,L}\partial_{x}(\log\sqrt{\rho})^{H}
    -(J_{R,L}\partial_{x}\log\sqrt{\rho})^{H}\right] =0.
 \la{xieq}
\eea 
Here $\rho$ is a function of $J_{R}$ and $J_{L}$ implicitly
given by (\ref{RL}). The Hamiltonian acquires a Sugawara-like form
\be
 \la{49}
    H=\frac{1}{8}\int dx\,   \rho \Big[(J_R+J_L)^2+(J_R^H-J_L^H)^2\Big]
\ee
with Poisson brackets
\bea
    \{J_{R,L}(x),J_{R,L}(y)\} &=& \pm 2\pi g\partial_x\delta(x-y)
 \la{P} \\
    & & \pm  \frac{g}{2L}\partial_x\partial_y
    \left[\left(\frac{1}{\rho(x)}+\frac{1}{\rho(y)}\right)\cot\frac{\pi}{L}(x-y)\right],
 \nonumber \\
    \{J_R(x),J_L(y)\} &=& -\frac{g}{2L}\partial_x\partial_y
    \left[\left(\frac{1}{\rho(x)}-\frac{1}{\rho(y)}\right)\cot\frac{\pi}{L}(x-y)\right].
\eea We note that Poisson brackets become canonical and left and
right fields decouple in the limit of a constant density.

\subsection{Chiral Reduction}
\la{sec:chiraldBO}

We first note that the right and left currents $J_{R,L}$ are not
separated in eq.~(\ref{xieq}). The equations for $J_{R}$ and $J_{L}$
are coupled through the density $\rho$ which should be found in
terms of $J_{R,L}$ from (\ref{RL}). However, it is possible to find
the \textit{chiral reductions} of 2BO assuming that one of the
currents is constant.  We explain this reduction in some detail
in this section.

The 2BO (\ref{2BO}) or (\ref{xieq}) admits an additional reduction to a chiral
sector \cite{2006-BAW-PRL-shocks} where one of the chiral currents
(\ref{RL}), say left current, is a constant $J_L(x,t)=v_0-\pi
g\rho_0$. We can always choose a coordinate system moving with
velocity $v_0$. This is equivalent to setting the zero mode of
velocity to zero $v_0=0$. The condition $J_{L}=-\pi g \rho_{0}$ becomes
\be
 \la{chiralconstr1}
    v=g\left[\pi  (\rho-\rho_0)+\partial_{x}(\log\sqrt\rho)^{H}\right].
\ee
Then the currents can be expressed in terms of the density field only
\bea
    J_{L}(x) &=& J_{0},
 \nonumber \\
    J_{R}(x) &=& J_{0} +J(x),
 \la{48} \\
    J_{0} &=& \pi g\rho_{0},
 \nonumber \\
    J(x) &=& 2g\left[\pi \left( \rho-\rho_0\right)+\partial_{x}(\log\sqrt\rho)^{H}\right].
\eea It follows from Eq.~(\ref{xieq}) that once the  current $J_{L}$
is  chosen to be constant $J_{L}(x)=J_{0}$ at $t=0$ it remains
constant at any later time. The condition (\ref{chiralconstr1}),
therefore, is compatible with 2BO. Then the   density $\rho(x,t)$
evolves according to  the continuity equation (\ref{continuity})
with velocity determined by the density according to
(\ref{chiralconstr1}). We obtain an important equation (written in
the coordinate system moving with velocity $v_0$) \be
    \rho_{t}+g\left[\rho\left(\pi\left (\rho-\rho_0\right)
    + \partial_{x}(\log\sqrt\rho)^{H}\right)\right]_{x} =0.
 \la{cheq10}
\ee We refer to this equation as the Non-Linear Chiral Equation
(NLC). A substitution  of the chiral constraint (\ref{48}) to
(\ref{49}) gives the  Hamiltonian  for NLC 
\bea
    H &=& \frac{1}{8}\int  dx\, \rho  \left[J^2+(J^H)^2\right]
 \label{JJ}
\eea
with Poisson brackets for $J(x)$ following from (\ref{P}). This equation constitutes one of major results of this paper.

NLC can be written in several useful forms. One of them is: \be
 \la{chi}
    \varphi_{t}+g\left[\pi \rho_0\left(2e^{\varphi}-\varphi\right)
    +\frac{1}{2}\varphi_{x}^{H}\right]_{x} +\frac{g}{2}\varphi_x \varphi^{H}_{x}=0,
\ee
where $\rho(x)=\rho_0\, e^{\varphi(x)}$.

\subsection{Holomorphic Chiral field}

Under the chiral condition (\ref{chiralconstr1}) the field $u_0$
becomes analytic  inside the disk. Indeed, combining (\ref
{chiralconstr1}) and (\ref{u+hydr}) we obtain \be
    u_0(w)=\frac{1}{2}\oint\frac{d\zeta}{2\pi i\zeta}\, \frac{\zeta+w}{\zeta-w}J(\zeta),\quad |w|<1.
\ee In the chiral case it  has only non-negative powers of $w$ in
the expansion (\ref{38}).  Negative modes vanish $b_n=0$.
Conversely, the condition of $u_0$ to be analytic inside the unit
disk is equivalent to $J_{L}=const$.

The current itself (\ref{48}) is the boundary value of the  field
$\R u_0$ harmonic  inside the disk \be
    J(x)=2J_{0}+2\R u_0 =2J_0+\sum_{n=1^{\infty}}\left(a_{n}w^n+\bar a_{n}w^{-n}\right).
\ee

The fields $u$  and $\tilde u$ are in turn also analytic inside the
disk. Let $\varphi$ to be a harmonic function inside the disk with
the boundary value $\log(\rho/\rho_{0})$. Then
$\varphi=\phi(w)+\overline{\phi(w)}$, where $\phi(w)=
(\log(\rho/\rho_{0}))^{+}$. Here $
f^{+}(w)=\frac{1}{2}\int\frac{\zeta+w}{\zeta-w}f(\zeta)\,\frac{d\zeta}{2\pi
i \zeta}$ is a function analytic in the interior of a unit circle
which value on the boundary of the disk is $(f(x)-if^{H}(x))/2$.  It
follows from (\ref{ux},\ref{chiralconstr1}) that \bea
    u=-J_{0}+ig \p \phi,\quad  |w|<1,
 \\
    \tilde u=u+4\pi g\rho_0 \left(e^{\varphi}\right)^{+},\quad  |w|<1,
\eea
Then 2BO (\ref{2BO1}) becomes an equation on an analytic function in the interior of a unit circle
\be
    \dot\phi+i\frac{g}{2} \left[(\p\phi)^2+\p^2\phi\right]
    +\pi g\rho_0\p\left(2e^{\varphi}-\varphi\right)^{+}
    =0.
\ee This is the ``positive part'' of (\ref{chi}) which is a direct
consequence of (\ref{cheq10}).

We remark here that the chiral equation (\ref{cheq10}) has a
geometric interpretation  as an evolution equation describing the
dynamics of a contour on a plane. Within this interpretation the
term $\partial_{x}(\log\sqrt{\rho})^{H}$ of (\ref{cheq10}) is the
curvature of the contour (see \ref{sec:geom}).

\subsection{Benjamin-Ono Equation}

Another form of the Chiral Equation (\ref{cheq10}) arises when one
considers the fields $u$ and $\tilde u$ outside the disk. There
neither $u$ nor $\tilde u$ are analytic, but their boundary values
are connected by the Hilbert transform \bea
 \la{12}
    u(x-i0) &=& -J_{0}+2g \left[\pi \rho +i\partial_{x}(\log\sqrt{\rho})^{+}\right],
 \\
    \tilde u(x-i0) &=&-J_{0} -i u^H(x-i0).
\eea 
The bidirectional equation Eq.~(\ref{2BO1}) complemented by
this condition becomes  unidirectional (chiral) 
\bea
 \la{BO2}
    &&u_{t}+\partial_{x}\left[\frac{1}{2}u^{2}+\frac{g}{2} \partial_{x} u^H \right] =0.
\eea
This is just another form of the chiral equation (\ref{cheq10}).

The chiral equation (\ref{BO2}) has the form of the
Benjamin-Ono equation \cite{Benjamin-Ono}. There are noticeable
differences, however. Contrary to the Benjamin-Ono equation,
Eq.~(\ref{BO2}) is written on a complex function, whose real and
imaginary values at real $x$ are related by conditions (\ref{12})
implementing the reality of the density: \be
 \la{C}
    \R u = -J_{0}+ 2g\pi\rho+g\left(\p_x\log\sqrt\rho\right)^H, \quad \I u=g\p_x\log\sqrt\rho.
\ee One understands this relation as a condition on the initial
data. Once it is imposed  by choosing the initial data for the
density $\rho$, the condition remains intact during the evolution.

However, in the case when the deviation of a density is small with
respect to the average density $|\rho-\rho_0|\ll \rho_0$, the
imaginary part of $u$ vanishes in the leading order of $1/\rho_0$
expansion
$$
    u\approx J_{0}+ 2\pi g\varphi\approx 2\pi g (\rho-\rho_0) +J_{0}
$$
and the condition (\ref{C}) becomes non-restrictive. In this limit
Eq.~(\ref{BO2}) becomes an equation on a  single real function. It
is the conventional Benjamin-Ono equation.
One can think of NLC (\ref{cheq10}) as of finite amplitude extension of BO. 
Similarly, 2BO is an integrable bidirectional finite amplitude extension of BO. It is interesting that there exists another bidirectional finite amplitude extension of BO -- the Choi-Camassa equation \cite{1996-ChoiCamassa}. However, it seems that the latter is not integrable.

\section{Multi-phase solution}
\la{sec:Multi-phase}

In this section we describe the most general finite dimensional
solutions of 2BO. These are multi-phase solutions and their
degenerations -- multi-soliton solutions. In the former case the
$\tau$-functions are polynomials of $e^{ik_ix}$, where $k_i$ is a
finite set of parameters, the latter are just polynomials of $x$.
These solutions are given by determinants of finite dimensional
matrices. They  appeared in the arXiv version  of
Ref.~\cite{2006-BAW-shocks}. One can construct those solutions
using the transformation (\ref{back1}) of 2BO to INLS (\ref{INLS}).
For the latter multi-phase solutions were written in \cite{1995-Pelinovsky} (see also \cite{2004-Matsuno,2004-Matsuno-INLS}). We use a different route in this section deriving multi-phase and multi-soliton solutions as a real reduction of corresponding solutions for MKP1.

\subsection{Multi-phase and multi-soliton solutions of MKP1}

We start from a general  multi-phase solution of MKP1 equation
and then restrict it to 2BO equation.

A general multi-phase solution of MKP1 equation
\bea
    \left(iD_{t}+\frac{g}{2}\, D_{x}^{2}\right)  \tau_{1}\cdot \tau_{0} =0.
 \la{MKP1}
\eea is given by the following determinant formulae
 \cite{1979-SatsumaIshimori,Matsuno-book}
\bea
    \tau_a &=& e^{i\theta_a} \det\left[\delta_{jk}
    +c_{a,j} \frac{e^{i\theta_{j}}}{p_{j}-q_{k}}\right],\quad a=0,1
 \la{86}\\
    \frac{c_{1,j}}{c_{0,j}}&=& \frac{q_{j}-K-v_0}{p_{j}-K-v_0},
 \la{relcc}
\eea
where the phases are
\bea
    g\,\theta_{j}(x,t) &=&
    (q_{j}-p_j)(x-x_{0j})-\frac{q_{j}^{2}-p_j^2}{2}t,
 \la{thetaj} \\
    g\, \theta_0(x,t) &=& Kx-\frac{K^{2}}{2}t
    -\left((v_{0}+K)x-\frac{(v_{0}+K)^{2}}{2}t\right),
 \la{thetap} \\
    g\, \theta_1(x,t) &=& Kx-\frac{K^{2}}{2}t.
 \la{thetam}
\eea This solution is characterized by an integer number $N$ (number
of ``phases''), and by $4 N-1$ parameters $p_{j}$, $q_{j}$,
$c_{0,j}$, $x_{0j}$ and  moduli $K$ and $v_0$. The solutions become
single-valued on a unit circle if  $p_{j}$ and $q_{j}$ are integers
in units of $g\frac{2\pi}{L}$.

\subsection{Multi-phase solution of 2BO}
\la{sec:mps2BO}

Without further restrictions the parameters entering
(\ref{86}-\ref{thetam}) are general complex numbers. Reality nature
of 2BO equation restricts them to be real.

The real moduli $K$ and $v_0$ are obviously zero modes of the fields
$u_1$ and $u_0$ respectively, and therefore, they are zero modes of
the density $\rho_{0}=\frac{1}{L}\int  \rho \,dx= -K/(\pi g)$ and
velocity $\frac{1}{L}\int  v \,dx= v_{0}$. 

\subsubsection{Schwarz reflection condition}
We have to restrict the coefficients $c_{a,j}$, so that there exists another
solution $\tau_{-1},\tau_0$ of Eq.~(\ref{MKP1}) sharing the same
$\tau_0$ with the solution  (\ref{MKP1sol-t},\ref{relcct}) and
obeying the Schwarz reflection property (\ref{SchwarzTau}).

The Galilean symmetry of the equation (\ref{MKP1}) is here to help.
If $\tau_a(x,t),\,a=0,1$ give a solution of (\ref{MKP1}) then  the
pair $e^{iP_ax-iE_at}\tau_a(x-gP_at,t)$  is also a solution provided
that $E_{a+1}-E_a=\frac{1}{2}(P_{a+1}-P_a)^2$.

Being applied to the solution (\ref{86}-\ref{thetam}) the Galilean invariance can be utilized as follows. We notice from (\ref{thetaj}) that
\be
    \theta_{j}(x-Pt, t; \left\{p_{j},q_{j}\right\}) = \theta_{j}(x, t; \left\{p_{j}+P,q_{j}+P\right\}).
 \la{shiftprop}
\ee
Performing the Galilean boost to (\ref{86}), multiplying both tau-functions by $e^{-i Pv_{0}t}$ and shifting $p_{j}\to p_{j}-P$, $q_{j}\to q_{j}-P$  we obtain that
\bea
    \tau_{-1} &=& e^{i\theta_1-i P (K+v_{0})t+\frac{i}{g}(Px-\frac{P^{2}}{2}t)} \det\left[\delta_{jk}
    +b_{j} \frac{e^{i\theta_{j}}}{p_{j}-q_{k}}\right],
 \la{MKP1sol-t} \\
    \frac{b_{j}}{c_{0,j}} &=& \frac{q_{j}-K-v_{0}-P}{p_{j}-K-v_{0}-P}
 \la{relcct}
\eea
form a solution of (\ref{MKP1}) with the same $\tau_0$ (\ref{86}).

Now we are going to show that for a particular choice of
coefficients $c_{1,j}$ the Galilean boosted solution
(\ref{MKP1sol-t},\ref{relcct}) is a complex conjugate of $\tau_{+1}$
from (\ref{86}). To show this we will employ the determinant
identity (\ref{detid}).\footnote{A similar trick was used by Matsuno
\cite{2004-Matsuno} to prove the reality of a multi-phase solution
for conventional Benjamin-Ono equation.}

We apply the determinant identity (\ref{detid}) to (\ref{MKP1sol-t}) and obtain
\bea
    {\tau}_{-1} &=&
    e^{i\theta_1-i P (K+v_{0})t+\frac{i}{g}(Px-\frac{P^{2}}{2}t)}
    \left(\prod_{j}
    e^{i\theta_{j}}\sqrt{\frac{b_{j}}{\tilde{b}_{j}}}\right) \det\left[\delta_{jk}
    +\tilde{b}_{j}
    \frac{e^{-i\theta_{j}}}{p_{j}-q_{k}}\right],
 \la{MKP1sol-trans}
\eea
where
\be
    \tilde{b}_{j} b_{j}= (p_{j}-q_{j})^{2}
    {\prod_{k \neq j}}\frac{(p_{j}-q_{k})(q_{j}-p_{k})}{(p_{j}-p_{k})(q_{j}-q_{k})}.
 \la{btjbj}
\ee
The Schwarz reflection condition(\ref{SchwarzTau})  ${\tau}_{-1}=\overline{\tau_{+1}}e^{i\Theta(t)}$  requires
\be
    \tilde{b}_{j} = c_{1,j},
 \la{btj}
\ee
determines   $ \Theta(t)$, and gives a relation
\be
    P=\sum_j(p_j-q_j)-2K.
 \la{PpqK}
\ee

Finally, combining all relations (\ref{relcc},\ref{relcct},\ref{btjbj},\ref{btj}) together we obtain  the condition on coefficients $c_{a,j}$
\bea
 \label{caj2BO}
    \left(\frac{c_{a,j}}{p_{j}-q_{j}}\right)^{2}
    &&{\prod_{k \neq j}}\frac{(p_{j}-p_{k})(q_{j}-q_{k})}{(p_{j}-q_{k})(q_{j}-p_{k})}
 \\
    &=&
    \left(\frac{p_{j}-K-v_{0}}{q_{j}-K-v_{0}}\right)^{1-2a}
    \frac{p_{j}-K-v_{0}-P}{q_{j}-K-v_{0}-P}.
 \nonumber
\eea

Condition (\ref{caj2BO}) is necessary to turn a general solution of MKP1 into a solution of 2BO. We also have to find a condition that guarantees that $\tau_{1}$ has no zeros inside the unit disk. Before  turning to this analysis, we  first discuss degeneration of formulas (\ref{86}) into multi-soliton solution.

\subsubsection{Multi-soliton solution of 2BO}
The multi-soliton solution of 2BO  follows  from the multi-phase solution in the limit $p_{j}\to q_{j}$. We introduce
\bea
    k_{j} &=& p_{j}-q_{j},
 \\
    v_{j} &=& \frac{1}{2}(p_{j}+q_{j})
\eea
and consider the limit $k_{j}\to 0$ keeping $v_{j}$ fixed. After some straightforward calculations we obtain
\bea
    \tau_a &=& e^{i\theta_a} \det\left[\delta_{jk} \left( x-x_{0j}-v_{j}t+iA_{a,j}\right)
    +i g \frac{1-\delta_{jk}}{v_{j}-v_{k}}\right],
 \la{taupmmsol}\\
    A_{a,j} &=& \frac{g}{2} \left(\frac{1}{v_{j}-v_{0}+K}\pm \frac{1}{v_{j}-v_{0}-K}\right),\quad a=0,1.
 \la{xij}
\eea 
One  notices that in the limit $t\to +\infty$ the  solution
(\ref{taupmmsol}) asymptotically goes to the factorized form 
\bea
    \tau_a \to e^{i\theta_a}\prod_{j}(x-x_{0j}-v_{j}t+iA_{a,j}),
 \la{taufact}
\eea describing separated single solitons.

Eq.~(\ref{taufact}) gives a large time value of zeros of $\tau_1$. Their imaginary part is
\be
    -\R A_{1,j} = g \frac{K}{(v_{j}-v_{0})^{2}-K^{2}}.
 \la{Ajmlim}
\ee
It must be negative in order for $\tau_1$ to have no zeroes inside the unit disk.  Since  $K<0$ we must require
\be
 \la{103}
    (v_j-v_0)^2>K^2.
\ee
In the next paragraph we argue that under this condition and additional restrictions on parameters $p_{j}$, $q_{j}$ (see eq.~(\ref{ordering1}) below) the moving
zeros never cross the real axis, and therefore zeros stay outside of
the unit disk at all times.

To conclude this section we note a unique property of the 2BO
equation (shared with the BO equation). Namely, there is a
``quantization'' of the mass of solitons: each soliton of 2BO
carries a unit of mass regardless of its velocity. We have for $N$-soliton solution
\be
    \int dx\, (\rho-\rho_{0})=N.
 \la{mscharge}
\ee
Where $K=-\pi g\rho_{0}$.
The total  momentum, and the total energy of a multisoliton solution is given by
\bea
    && \int dx\, (\rho v-\rho_{0}v_{0}) = \sum_{j}v_{j},
 \la{msmom}\\
    && \int dx\,\left(\frac{\rho v^{2}}{2}+\rho\epsilon(\rho)
    -\frac{\rho_{0} v_{0}^{2}}{2}-\rho_{0}\epsilon(\rho_{0}) \right)
    = \sum_{j}\frac{v_{j}^{2}}{2},
 \la{msenergy}
\eea
where $\epsilon(\rho)$ is defined in (\ref{epsrho0}).

One-soliton solution has a form
\bea
    \rho &=& \rho_{0} +\frac{1}{\pi} \frac{A_{1}}{(x-x_{01}-v_{1}t)^{2}+A_{1}^{2}},
 \\
    v &=& v_{0} +g \frac{A_{0}}{(x-x_{01}-v_{1}t)^{2}+A_{0}^{2}},
\eea
where
\bea
    A_{1} &=& \R A_{1,1} = \frac{\pi g^{2}\rho_{0}}{(v_{j}-v_{0})^{2}-(\pi g\rho_{0})^{2}},
 \\
    A_{0} &=& \R A_{0,1}= \frac{g(v_{j}-v_{0})}{(v_{j}-v_{0})^{2}-(\pi g\rho_{0})^{2}}.
\eea
This one-soliton solution has been found first in Ref.~\cite{1995-Polychronakos} (see also \cite{AndricBardekJonke-1995}).

\subsubsection{Analyticity condition}

Now we can turn to the multiphase solution and derive conditions
sufficient in order for $u_{1}$ to be analytic in the upper
half-plane in complex $x$-variable (inside the unit disk).
Analyticity in the lower half-plane follows from the Schwarz
reflection condition (\ref{SchwarzTau}).  We will follow the
approach of Dobrokhotov and Krichever
\cite{1991-DobrokhotovKrichever} developed for Benjamin-Ono
equation.

Analyticity  of $u_1$ means that  $\tau_1$  given by (\ref{86}) has no zeros in the upper half plane, or that the  matrix
\bea
    M_{jk} =\delta_{jk}
    +\frac{c_{1,j}
    e^{i\theta_j} }{p_{j}-q_{k}}
 \la{M-matrix}
\eea
is non-degenerate.
Following the approach of Ref. \cite{1991-DobrokhotovKrichever} we
derived in \ref{sec:ndc} a sufficient condition of non-degeneracy of
the matrix $M$ from (\ref{M-matrix}). Let us now write that
condition (\ref{mndcond},\ref{fj}) with $c_{j}$ defined by
(\ref{cj1},\ref{caj2BO}). We obtain (calculating $f_{j}$) 
\be
    \frac{P}{\tilde{p}_{j}(\tilde{q}_{j}-P)}
    \prod_{k(k\neq j)}\frac{\tilde{p}_{j}-\tilde{q}_{k}}{\tilde{p}_{j}-\tilde{p}_{k}}
     \;\; \mbox{same sign for all }j,
 \la{f1cond}
\ee
where we used shifted numbers $\tilde{p}_{j}=p_{j}-K-v_{0}$ and similar for $\tilde{q}_{j}$.
Here $P=-2K +\sum_j(\tilde{p}_j-\tilde{q}_j)$.

The set of conditions 
\bea
    \tilde{q}_{1} <\tilde{p}_{1}< \ldots <\tilde{q}_{m} <\tilde{p}_{m}
    <0 <P
    < \tilde{q}_{m+1} <\tilde{p}_{m+1}< \ldots <\tilde{q}_{N} <\tilde{p}_{N}
 \la{ordering1}
\eea 
satisfies  (\ref{f1cond}).  Moreover, (\ref{ordering1})  yields
to (\ref{103}), which in its turn means that at least at some values
of parameters (large time and soliton limit) no zeros of $\tau_1$
are  inside the unit disk. Since they also can not be on the circle
they do not cross it while moving in time and in the space of
parameters.

Condition (\ref{ordering1}) suggests that a general solution is
characterized by a integer number $N-2m$. This is chirality -- the
difference between the number of $N-m$ right  and $m$ - left moving
modes 
\bea
    \frac{1}{2\pi g}\int (J_R-v_{0}-\pi g\rho_{0})\,dx
    &=& N-m,
 \\
    \frac{1}{2\pi g}\int (J_L-v_{0}+\pi g\rho_{0})\,dx
    &=& m.
\eea 
Eqs.~(\ref{86}-\ref{thetam},\ref{caj2BO},\ref{ordering1})
summarize a general finite dimensional quasi-periodic solution. We 
emphasize here that this solution is not chiral and contains both right and left-moving modes.

\subsection{Multi-phase solution of the Chiral Non-linear Equation}

The (right) chiral case appears when $\tau_0$  has no zeros outside the unit disk. It naturally  happens when the number of, say, left-moving modes $m$ in (\ref{ordering1}) vanishes $m=0$.  In this case all $v_{j}-v_{0}<0$. In their turn,  imaginary parts of zeros of $\tau_0$ in the  multi-soliton limit (as in (\ref{Ajmlim}))
\be
    -A_{0j} =- g \frac{v_{j}-v_{0}}{(v_{j}-v_{0})^{2}-K^{2}}>0
 \la{AJplim}
\ee are positive. One can check that in this case (\ref{M-matrix}) with $c_{1,j}\to c_{0,j}$ is non-degenerate for arbitrary values of parameters satisfying (\ref{ordering1}) with $m=0$ (and similarly for  $m=N$). 
Therefore, $\tau_{0}(x)$ is non-zero in one
of half-planes.

This is a chiral multi-phase solution of 2BO.

\subsection{Multi-phase solution of the Benjamin-Ono equation}

The known solutions of the Benjamin-Ono equation \cite{CLP-1979,1979-SatsumaIshimori} are obtained from the solutions of the Chiral Non-linear equation by taking the limit $\rho_{0} \to \infty$. In this case $K\to-\infty$ and conditions (\ref{ordering1}) allow for a good limit only if $m=N$ (left sector) or $m=0$ (right sector). Let us concentrate on the right sector. We redefine $p_{j}\to p_{j}-K$, $q_{j}\to q_{j}-K$, $v_{0}\to v_{0}-2 K$, go to the frame moving with velocity $-K$ ($x\to x+Kt$), and obtain from (\ref{ordering1},\ref{caj2BO}) in the limit $K\to -\infty$
\bea
    q_{1}<p_{1}<q_{2}<\ldots<q_{N}<p_{N},
 \label{ordering2}
\\
    \left(\frac{c_{a,j}}{p_{j}-q_{j}}\right)^{2}
    {\prod_{k \neq j}}\frac{(p_{j}-p_{k})(q_{j}-q_{k})}{(p_{j}-q_{k})(q_{j}-p_{k})}
    =
    \left(\frac{p_{j}-v_{0}}{q_{j}-v_{0}}\right)^{1-2a}
 \label{cajBO}
\eea
with solution given by (\ref{86},\ref{thetaj}) and with (\ref{thetap},\ref{thetam}) (one should put $K=0$ in latter two). This is nothing else but the multiphase solution of conventional Benjamin-Ono equation \cite{1979-SatsumaIshimori,2004-Matsuno}.

\subsection{Moving Poles}

The 2BO equation (\ref{2BO}) looks very similar to the classical BO
equation. One of important tools in studying the classical BO
equation is the so-called pole ansatz -  solutions in the form of
poles moving in a complex plane. \cite{CLP-1979} We have already seen that the pole
ansatz (\ref{u-pa},\ref{u+pa}) describes the dynamics of the
original Calogero-Sutherland model with finite number of particles
$N$.

In this section we consider collective excitations of
Calogero-Sutherland model in the limit of infinitely many particles.
These excitations are given by ``complex'' pole solutions of the
2BO.

In the Pole Ansatz (\ref{u-pa},\ref{u+pa}), the reality conditions
were satisfied by requiring $x_{j}$ to be real (or $w_{j}(t)$ moving
on a unit circle). One could generalize the Pole Ansatz
(\ref{u-pa},\ref{u+pa}) to case where $w_{j}(t)$ are away from the
unit circle and moving in a complex plane. The equations
(\ref{pmotx},\ref{pmoty}) describing the motion of poles preserve
their form. However, $u_{-1}(w)$ outside of the unit circle is not
related to the $u_{1}(w)$ inside of the circle by analytic
continuation but only by Schwarz reflection (\ref{Schwarz}). The
field $u_{1}(w)$ is analytic inside the unit circle and has moving
poles outside of the unit circle (and vice versa for $u_{-1}(w)$).
Of course, having obtained the solution of 2BO inside the unit
circle does not mean automatically that the Schwarz reflected
function (\ref{Schwarz}) will solve 2BO in the exterior of the
circle \textit{with the same} $u_0$. The property (\ref{Schwarz})
requires that (\ref{pmotx},\ref{pmoty}) are satisfied not only by
$u_{j}$ and $w_{j}$ but also by $u_{j}$ and $1/\bar{w}_{j}$. This
requirement will significantly constrain the positions of poles
$w_{j}$ and $u_{j}$ in a complex plane. It turns out 
that this constraint allows for non-trivial solutions.

We emphasize here once again that while real axis poles $x_{j}$ of
$u_{1}$ in the pole ansatz represent the original CS particles, the
complex poles $x_{j}$  represent collective excitations of the CS
liquid moving in the background of macroscopic number of particles.

Instead of looking for moving pole solution in this section we have
taken a different route. We first construct the much more general
solution of 2BO (\ref{2BO}) with proper reality conditions and then
obtain a moving pole (i.e., multi-soliton) solution as a limit of
the multi-phase solution. One can see from (\ref{taupmmsol}) that for soliton solutions
the zeros of tau-functions move in a complex plane. It is especially clear at large times 
when solitons are well separated (\ref{taufact}).

\section{Conclusion and discussion}

In this paper we have shown that the dynamics of the classical
Calogero-Sutherland model in the limit of infinite number of
particle is equivalent to the bidirectional Benjamin-Ono equation
(\ref{2BO}). The bidirectional Benjamin-Ono equation (2BO) is an
integrable classical integro-differential equation. Its
integrability can be deduced from the fact that it is a Hamiltonian
reduction of MKP1 as it is shown in this paper. As an alternative,
one can use the equivalence of 2BO to INLS  (\ref{INLS}). The
integrability of INLS  was proven and the spectral transform was
constructed for INLS  in Ref. \cite{1995-PelinovskyGrimshaw} (see
also \cite{2004-Matsuno-INLS}). Therefore, one can use all techniques
developed in the field of classical integrable equations for 2BO. It
has multi-phase solutions (explicitly constructed in this paper),
bi-Hamiltonian structure, an associated hierarchy of higher order
equations, etc. 2BO is intrinsically simpler than many other
classical integrable models. Its solitons have ``quantized'' area
independent of soliton's velocity. The collision of two solitons
goes without any time delay etc. This is a reflection of the fact
that underlying Calogero-Sutherland model is essentially a model of
non-interacting particles in disguise. In particular, 2BO supports a
phenomenon of dispersive shock waves. Some applications of this
phenomenon to quasi-classical description of quantum systems were
considered in \cite{2006-BAW-PRL-shocks}.

Most of the results of this paper can be generalized along two
avenues: generalization to an elliptic case and generalization to a
quantum model.

The Calogero-Sutherland model (trigonometric case) can be
generalized to an elliptic case --- elliptic Calogero model where
the interaction between particles  is either Weierstrass
$\wp(x|\omega_1,\omega_2)$-function with purely real and purely
imaginary periods $\omega_1,i\omega_2$, and to its hyperbolic
degeneration (hyperbolic case) with inter-particle interaction given
by $\sinh^{-2}(x/\omega_2)$ (see \cite{1981-OlshanetskyPerelomov-classical} for review).

In both cases most of formulas remain unchanged if one substitutes
the Hilbert transform $f^H$ for a  transform with respect to a strip
$0<\I x<\omega$, where $\omega$ is an imaginary period 
\be
 \la{Helliptic}
    f^H=\int \zeta(x-x')f(x')dx'\;\;\;\; \mbox{or} \;\;\;
    \int \frac{1}{\omega_2} \coth\frac{1}{\omega_2}(x-x')f(x')dx'.
\ee In the first case the integration goes over a real period of the
Weierstrass $\zeta$-function.

The elliptic  Calogero model  allows one to study a crossover
between liquids with long range inter-particle interaction to
liquids with short range interaction. In the limit of a large
imaginary period  $\omega_{2}\to\infty$ the $\wp$-function
degenerates to $1/\sin(x/\omega_1)^2$ -- the case of long range
inter-particle interaction. The opposite limit  $\omega_2\to 0$
gives rise to a short range interaction: $\omega_2\wp(x)\to
\delta(x)$.

In the latter case the the Hilbert transform (\ref{Helliptic}) becomes a
derivative $f^H\to \omega\p_x f$ and the equations discussed  in
this paper become local. In particular the Benjamin-Ono equation
flows to the KdV equation, while the bidirectional BO-equation flows
to NLS - the nonlinear Schr\"odinger equation.

2BO in the limit of small amplitudes and in the chiral sector becomes
the conventional Benjamin-Ono equation. In elliptic
case (and in the hyperbolic one) the limit of small amplitudes in
the chiral sector leads to a generalization of the Benjamin-Ono
equation, known as the ILW (Intermediate Long Wave)  equation
\cite{AblowitzClarkson-book}. Contrary to the Benjamin-Ono equation
and to 2BO, the latter and its bidirectional generalization
2ILW have elliptic solutions.

We intend to address the elliptic case in a separate publication.

Probably, even more interesting is a generalization of the results of this paper
to the quantum case. It is well known that the classical CSM model
(\ref{CSM}) can be lifted to a quantum integrable
Calogero-Sutherland model \cite{Calogero-1969,Sutherland-1971,1999-Polychronakos}. The
latter model is defined by (\ref{CSM}) with $g^2 \to
\hbar^2\lambda(\lambda-1)$ and $p_i=-i\hbar\p_{x_i}$. The 2BO
equation in the form (\ref{2BO}) remains unchanged, except for the
change of the coefficient $g\to\lambda-1$ and for the change of
Poisson brackets (\ref{PB01}) by a commutator: $\{\,,\,\}\to
\frac{i}{\hbar}[\,,\,]$. The change $g\to \lambda-1$ valid for eq.
(\ref{2BO}) is not correct for all formulas. For example, 
the bilinear form of classical 2BO (\ref{2BO-hirota}) is identical
to its quantum version with just a change of notations $g\to
\lambda$.   For some details see \cite{2005-AbanovWiegmann}. 
Multi-soliton solution of 2BO presented here corresponds to exact 
quasiparticle excitations of quantum Calogero-Sutherland model \cite{Sutherland-book,1995-Polychronakos}. The
more detailed study of the relations between integrable structures
of the classical 2BO and its quantum analogue is necessary.

\section{Acknowledgments}

AGA is grateful to A. Polychronakos for the discussion of the chiral case. PW thanks J. Shiraishi for discussions.
The work of AGA was supported by the NSF under the grant DMR-0348358. EB was supported by ISF grant number 206/07. PW was supported by NSF under the grant  NSF DMR-0540811/FAS 5-27837 and MRSEC DMR-0213745. We also thank the Galileo Galilei Institute for Theoretical Physics for the hospitality and the INFN for partial support during the completion of this work.

\appendix

\section[\hspace{2.5cm}Hilbert transforms]{Hilbert transforms}
\la{app-hilbert}

Given a function $f(x)$, $f(x)\to 0$ as $x\to\pm \infty$, the Hilbert transform is defined as
\be
    f^{H}(x) =\frac{1}{\pi}\dashint_{-\infty}^{+\infty} dy\, \frac{f(y)}{y-x}.
  \la{Htransline}
\ee
For periodic functions with period $L$ we define the transform as
\be
    f^{H}(x) =\dashint_{0}^{L}\frac{dy}{L}\, f(y)\cot\frac{\pi}{L}(y-x).
 \la{HtransCapp}
\ee
The Hilbert transform of the constant function is zero.
The Hilbert transform is  inverse to itself
$H^{2} =-1$ or $(f^{H})^{H} =-f$
and it commutes with derivative
$(f^{H})_{x} =(f_{x})^{H}.$
More generally,  a function $f(x)$ defined on the closed (directed) contour $C$ surrounded the origin of a complex plane can be  decomposed into a sum  $f=f^{+}+f^{-}$,   analytic  functions $f^{\pm}$ inside (outside) of the contour, such that $f^+(0)=f^-(\infty)$. Then
\be
    f^{H} \equiv P\oint \frac{d\zeta}{2\pi\zeta}\, f(\zeta) \frac{\zeta+w}{\zeta-w}=i(f^{+}-f^{-}).
 \la{HtransGen}
 \ee
Using (\ref{HtransGen}) it is easy to derive the following properties:
\bea
    f^{H}g +f g^{H} &=& (fg)^{H} - (f^{H}g^{H})^{H}
\eea
and some integration formulas
\bea
    \int dx\, f^{H}  &=& 0,
 \\
    \int dx\, f^{H}g &=& -\int dx\, f g^{H},
 \\
    \int dx\, f^{H}f &=& 0.
\eea
From (\ref{HtransGen}) we have for functions analytic in one of half planes
\be
	(f^{\pm})^{H} =\pm i f^{\pm}
\ee
and as an immediate consequence
\bea
	(e^{ikx})^{H} &=& ie^{ikx}\sgn k,
 \\
    \left(\frac{1}{x-a}\right)^{H} &=&  \frac{i}{x-a} \;\;\;\; \mbox{for}\; \I a>0.
\eea
Generally, in Fourier space the Hilbert transform is equivalent to a multiplication by $i\sgn k$, i.e., for Fourier coefficients
\be
    (f^{H})_{k} = i(\sgn k)f_{k}.
\ee
It is easy to derive the following useful identities
\bea
    \int u^{2}\, dx &=& \int (u^{H})^{2}\, dx,
 \\
    \int u^{3}\, dx &=& 3\int u (u^{H})^{2}\, dx,
 \la{hid1} \\
    \int u^{4}\, dx &=& \int \left[(u^{H})^{4}+6u^{2}(u^{H})^{2}\right]\, dx,
 \\
    \int u^{5}\, dx &=& 5\int \left[u(u^{H})^{4}-2u^{3}(u^{H})^{2}\right]\, dx.
\eea

\section[\hspace{2.5cm}Conserved integrals of 2BO]{Conserved integrals of 2BO}
\label{app:int}

In this section we present the conserved integrals of 2BO written in different forms.
The contour integrals below are taken along the contour $C$ defined in Figure \ref{fig:leftright}.
\bea
    I_{1} &=&  \oint \frac{dx}{2\pi g}\, u
    =\int dx\,\rho = \int \frac{dx}{2\pi}\,i\partial_{x}\log\frac{\tau_{1}}{\tau_{-1}},
 \la{I1app} \\
    I_{2} &=&  \oint \frac{dx}{2\pi g}\, \frac{1}{2}u^{2}
    = \oint \frac{dx}{2\pi g}\, u_1u_0
    =\int dx\, \rho v
 \nonumber \\
    &=&\frac{g}{2}\int \frac{dx}{2\pi}\,
    \left[\frac{D_{x}^{2}\tau_{-1}\cdot\tau_{0}}{\tau_{-1}\tau_{0}}
    -\frac{D_{x}^{2}\tau_{1}\cdot\tau_{0}}{\tau_{1}\tau_{0}}\right]
    =i\partial_{t}\int\frac{dx}{2\pi}\,\log\frac{\tau_{1}}{\tau_{-1}},
 \la{I2app} \\
    I_{3} &=&  \oint \frac{dx}{2\pi g}\, \left[\frac{1}{3}u^{3}
    +i\frac{g}{2} u\partial_{x}\tilde{u}\right]
 \nonumber \\
    &=& \oint \frac{dx}{2\pi g}\,
    \left[{u_1}^{2}u_0+u_1{u_0}^{2}
    +i g u_1 \partial u_0\right]
    =\int dx\,\left[\frac{\rho v^{2}}{2}+\rho\epsilon(\rho)\right].
 \la{I3app}
\eea
Higher integrals of motion for 2BO can be constructed recurrently similarly to Benjamin-Ono equation \cite{1983-Matsuno} or can be written using the integrals obtained for INLS \cite{1995-PelinovskyGrimshaw,2004-Matsuno-INLS}.

\section[\hspace{2.5cm}Geometrical interpretation of the chiral equation]{Geometrical interpretation of the chiral equation: Contour dynamics}
\label{sec:geom}

Equation (\ref{cheq10}) can be cast in the form of contour dynamics.

Let us interpret the unit disk as a uniformization of a simply-connected domain embedded into the complex $z$-plane. In other words, $z(w)$ is a conformal map of the interior of the unit disk $|w|=1$ to a bounded  domain  such that the length element of its boundary is proportional to the density
\be
    ds\equiv |z'(x)|dx=\pi \rho(x)dx.
\ee
(equivalently $1/\sqrt{\rho}$ is a harmonic measure of the contour).
Then the equation (\ref{cheq10}) describes the evolution of the planar domain. We notice that the curvature of the boundary $\kappa=-i\p_s\log z_{s}$ can be expressed  in terms of the density $\rho(x)$ as
\be
    \kappa=-(\pi\rho)^{-1}\left(\p_x\left(\log\sqrt\rho\right)^H-1\right).
\ee
Then  (\ref{cheq10}) can be written as
\be
    \frac{ds}{dt} = g \left(\frac{ds}{dx} \right)^2(1-\kappa),
 \la{cdynamics}
\ee
where ${d}/{dt}=\p_t-g\pi \rho_0\,\p_x$ and the time derivative is taken at fixed $x$.
The equation (\ref{cdynamics}) describes the evolution of a planar contour driven by its curvature.

\section[\hspace{2.5cm}Determinant identities]{Determinant identity}
\label{sec:det}

Here we present a derivation of a determinant identity (\ref{detid}) which was used in Sec.~\ref{sec:mps2BO}.

Consider  the Cauchy  matrix
\be
    D_{ij} = \frac{r_{i}s_{j}}{p_{i}-q_{j}},\quad p_{i}\neq q_{j}, \qquad i,j=1,2,\ldots,N.
 \la{Clmatrix}
\ee
Its determinant
\be
    \det(D) = \prod_{i}r_{i}
    \frac{\prod_{i<j}(p_{i}-p_{j})(q_{j}-q_{i})}{\prod_{i,j}(x_{i}-y_{j})}\prod_{j}s_{j}
    =\left(\prod_{i}r_{i}s_{i}\tilde{r}_{i}\tilde{s}_{i}\right)^{-\frac{1}{2}},
 \la{detCl}
\ee
where $\tilde{r}_{i}$ and $\tilde{s}_{i}$ are  defined by
\bea
    \tilde{r}_{i} r_{i} &=& (p_{i}-q_{i}) {\prod_{k (k\neq i)}} \frac{p_{i}-q_{k}}{p_{i}-p_{k}},
 \la{rtilde}\\
    \tilde{s}_{i} s_{i} &=& -(q_{i}-p_{i}) {\prod_{k (k\neq i)}} \frac{q_{i}-p_{k}}{q_{i}-q_{k}}.
 \la{stilde}
\eea
The inverse of $D$ is also Cauchy matrix
(see e.g., \cite{1959-Schechter}) given by
\be
    (D^{-1})_{ij} = \frac{\tilde{r}_{j}\tilde{s}_{i}}{p_{j}-q_{i}}.
 \la{invCl}
\ee
    The obvious identity
\be
    \det(1+D) = \det(D) \det(1+(D^{T})^{-1})
\ee
being specialized  for Cauchy matrix $D$ (\ref{Clmatrix})  reads
\be
    \frac{\det\Big(\delta_{ij}+\frac{r_{i}s_{j}}{p_{i}-q_{j}}\Big)}
    {\left(\prod_{i}r_{i}s_{i}\right)^{1/2}}
    =
    \frac{\det\Big(\delta_{ij}+\frac{\tilde{r}_{i}\tilde{s}_{j}}{p_{i}-q_{j}}\Big)}
    {\left(\prod_{i}\tilde{r}_{i}\tilde{s}_{i}\right)^{1/2}}.
 \la{detid}
\ee

\section[\hspace{2.5cm}Non-degeneracy condition of matrix (\ref{M-matrix})]{Non-degeneracy condition of matrix (\ref{M-matrix})}
\la{sec:ndc}

In this section we will derive the non-degeneracy condition for the matrix (\ref{M-matrix}) following \cite{1991-DobrokhotovKrichever}.

Let us introduce
\be
    c_{j}=c_{1,j}\exp\left\{\frac{i}{g}\left[(q_{j}-p_{j})x-\frac{q_{j}^{2}-p_{j}^{2}}{2}t\right]\right\}.
\la{cj1}
\ee
Our aim is to derive a sufficient condition on coefficients $c_{j}$ for the matrix
\be
    M_{jk} =\delta_{jk}+\frac{c_{j}}{p_{j}-q_{k}}
 \la{M-matrix2}
\ee
to be non-degenerate for real $p_{j}$ and $q_{j}$.

Let us assume that the matrix (\ref{M-matrix2}) is degenerate. It means the the following equation has a non-zero solution
\bea
    r_{j} +\sum_{k}\frac{c_{j}}{p_{j}-q_{k}}r_{k} =0.
 \la{nondcond}
\eea
We introduce the meromorphic function
\be
    \psi(z) = \sum_{k}\frac{r_{k}}{z-q_{k}}
\ee
and rewrite the condition (\ref{nondcond}) as
\be
    r_{j}=\mbox{\rm res}_{z=q_{j}}\psi(z) = -c_{j}\psi(p_{j}).
\ee
Now, let us consider the function
\be
    f(z)  = \psi(z) \overline{\psi(\bar{z})} \prod_{j}\frac{z-q_{j}}{z-p_{j}}.
\ee
Here explicitly
\be
    \overline{\psi(\bar{z})} = \sum_{k}\frac{\bar{r}_{k}}{z-q_{k}}.
\ee
The function $f(z)$ is meromorphic and its residue at infinity is zero (as $\psi(z)\sim z^{-1}$ as $z\to \infty$). On the other hand ($p_{j}$ and $q_{j}$ are real numbers)
\bea
    &&\mbox{\rm res}_{z=p_{j}}f + \mbox{\rm res}_{z=q_{j}}f
    =
     \psi(p_{j}) \overline{\psi(p_{j})} (p_{j}-q_{j})\prod_{k(k\neq j)}\frac{p_{j}-q_{k}}{p_{j}-p_{k}}
 \nonumber \\
     &&+ r_{j}\bar{r}_{j} \frac{1}{q_{j}-p_{j}}\prod_{k(k\neq j)}\frac{q_{j}-q_{k}}{q_{j}-p_{k}}
    = \frac{|r_{j}|^{2}}{|c_{j}|^{2}}\,  f_{j},
 \la{condder}
\eea
where we introduced a notation
\be
    f_{j}=(p_{j}-q_{j})\prod_{k(k\neq j)}\frac{p_{j}-q_{k}}{p_{j}-p_{k}}
    \left[1-\frac{|c_{j}|^{2}}{(p_{j}-q_{j})^{2}}\prod_{k(k\neq j)}
    \frac{(p_{j}-p_{k})(q_{j}-q_{k})}{(p_{j}-q_{k})(q_{j}-p_{k})}\right].
 \la{fj}
\ee
If $f_{j}$ has the same sign for all $j$, e.g., $f_{j}<0$  we immediately obtain
$$
    \sum_{j}\left(\mbox{\rm res}_{z=p_{j}}f + \mbox{\rm res}_{z=q_{j}}f \right) <0.
$$
This is impossible as the residue of $f(z)$ at infinity is zero. This contradiction shows that the matrix (\ref{M-matrix2}) is nondegenerate under these conditions.

To summarize, the sufficient condition of non-degeneracy of (\ref{M-matrix2}) is
\be
    f_{j}<0 \;\; \mbox{for all }j \qquad \qquad
    \mbox{or } \qquad\qquad f_{j}>0 \;\; \mbox{for all }j,
 \la{mndcond}
\ee
where $f_{j}$ are defined according to (\ref{fj}).



\section*{References}



\begin{thebibliography}{99}

\bibitem{Calogero-1969}
    F. Calogero, J. Math. Phys. \textbf{10}, 2191, 2197 (1969); \textbf{12}, 419 (1971).

\bibitem{Sutherland-1971}
    B. Sutherland, Phys. Rev. A \textbf{4}, 2019 (1971); A \textbf{5}, 1372 (1972).
    Phys. Rev. Lett. \textbf{34}, 1083 (1975).

\bibitem{1981-OlshanetskyPerelomov-classical}
M. A. Olshanetsky and A. M. Perelomov,  Phys. Rep. \textbf{71},  pp. 313-400, (1981).
    \\ \textit{Classical integrable finite-dimensional systems related to Lie algebras.}

\bibitem{2005-AbanovWiegmann}
    A. G. Abanov and P. B. Wiegmann, Phys. Rev. Lett {\bf 95}, 076402 (2005).
 \\ \textit{Quantum Hydrodynamics, the Quantum Benjamin-Ono equation, and the Calogero Model.}

\bibitem{AJL-1983}
    I. Andri\'c, A. Jevicki, and H. Levine, Nucl. Phys. \textbf{B215},
   307 (1983).
  \\ {\it On the large-N limit in symplectic matrix models.}

\bibitem{AndricBardek-1988}
    I. Andri\'c and V. Bardek, J. Phys. A \textbf{21}, 2847 (1988).
   \\ {\it 1/N corrections in Calogero-type models
    using the collective-field method.}
 \\
        \textit{ibid}. \textbf{24}, 353 (1991).
 \\ {\it Collective-field method for a U(N)-invariant model
in the large-N limit.}

\bibitem{1995-Polychronakos}
    A. P. Polychronakos, Phys. Rev. Lett. \textbf{74}, 5153 (1995).
    \\ \textit{Waves and Solitons in the Continuum Limit of the Calogero-Sutherland Model.}

\bibitem{AblowitzClarkson-book}
    M. A. Ablowitz and P. A. Clarkson, \textit{Solitons, Nonlinear
    Evolution Equations and Inverse Scattering}, London Math. Society
    Lecture Note Series (No. 149), 1991.

\bibitem{Awata}
    H. Awata, Y. Matsuo, S. Odake, J. Shiraishi,
    Phys. Lett. \textbf{B 347}, 49-55, (1995).
    \\{\it Collective field theory, Calogero-Sutherland model
and generalized matrix models.}

\bibitem{JevickiSakita}
      A. Jevicki and B. Sakita, Nucl. Phys. \textbf{B165},
   511 (1980).
   \\ \textit{The Quantum Collective Field Method and its
   Application to the Planar Limit.}

\bibitem{Sakita-book}
    B. Sakita, \textit{Quantum Theory of Many-variable Systems and
Fields.} World  Scientific, 1985.

\bibitem{Jevicki-1992}
      A. Jevicki, Nucl. Phys. \textbf{B376}, 75-98 (1992).
   \\ \textit{Nonperturbative Collective Field Theory.}

\bibitem{Benjamin-Ono}
    T. B. Benjamin, J. Fluid Mech. \textbf{29}, 559 (1967).
 \\ {\it  Internal waves of permanent form in fluids of great depth.}
 \\
    H. Ono, J. Phys. Soc. Japan \textbf{39}, 1082 (1975).
 \\ {\it Algebraic solitary waves in stratified fluids.}

\bibitem{1995-Pelinovsky}
    D. Pelinovsky, Phys. Lett. \textbf{A 197}, 401-406 (1995).
 \\ \textit{Intermediate nonlinear Schr\"odinger equation for internal waves in a fluid of finite depth.}

\bibitem{2006-BAW-PRL-shocks}
    E. Bettelheim, A. G. Abanov, and P. Wiegmann,  Phys. Rev. Lett.  \textbf{97},  246401 (2006).
 \\ \textit{Nonlinear quantum shock waves in fractional quantum hall edge states.}

\bibitem{1996-ChoiCamassa}
    W. Choi and R. Camassa,  Phys. Rev. Lett. \textbf{77}, 1759-1762 (1996).
 \\ \textit{Long internal waves of finite amplitude.}

 \bibitem{2006-BAW-shocks}
    E. Bettelheim, A. G. Abanov, and P.  Wiegmann, appendix of arXiv:cond-mat/0606778 (2006), unpublished.
    \\ {\it Quantum Shock Waves - the case for non-linear effects in dynamics of electronic liquids.}

\bibitem{2004-Matsuno}
    Y. Matsuno, J. Phys. Soc. Jap., \textbf{73}, 3285-3293 (2004).
    \\ \textit{New Representations of Multiperiodic and Multi-soliton Solutions for a Class of Nonlocal Soliton Equations.}

\bibitem{2004-Matsuno-INLS}
    Y. Matsuno, Inv. Probl. \textbf{20}, 437-445 (2004).
 \\ \textit{A Cauchy problem for the nonlocal nonlinear Schr\"odinger equation.}

\bibitem{CLP-1979}
    H. H. Chen, Y. C. Lee, and N. R. Pereira, J. Phys. Fluids
\textbf{22}, 187 (1979).
   \\ \textit{Algebraic internal wave solitons
    and the integrable Calogero-Moser-Sutherland $N$-body problem.}

\bibitem{1979-SatsumaIshimori}
    J. Satsuma and Y. Ishimori, J. Phys. Soc. Jpn.,  \textbf{46}, pp. 681-687 (1979).
    \\ \textit{Periodic Wave and Rational Soliton Solutions of the Benjamin-Ono Equation.}

\bibitem{Matsuno-book}
    Y. Matsuno, \textit{Bilinear
Transformation Method}, in v. \textbf{174}, Math. in science
and engineering, Academic Press, 1984.

\bibitem{AndricBardekJonke-1995}
    I. Andri\'c, V. Bardek, L. Jonke, Phys. Lett. \textbf{B 357},
   374 (1995).
   \\ {\it Solitons in the Calogero-Sutherland collective-field model.}

\bibitem{1991-DobrokhotovKrichever}
    S. Yu. Dobrokhotov and I. M. Krichever, Math. Notes. \textbf{49}, 583-594 (1991).
    \\ \textit{Multi-phase solutions of the Benjamin-Ono equation and their averaging.}

\bibitem{1995-PelinovskyGrimshaw}
    D. E. Pelinovsky and R. H. J. Grimshaw, J. Math. Phys. \textbf{36}, 4203-19 (1995).
 \\ \textit{A spectral transform for the intermediate nonlinear Schr\"odinger equation.}
 
\bibitem{1983-Matsuno}
    Y. Matsuno, J. Phys. Soc. Jap., \textbf{73}, 2955-2958 (1983).
    \\ \textit{Recurrence Formula and Conserved Quantity of the Benjamin-Ono Equation.}

\bibitem{1999-Polychronakos}
    M. A. Olshanetsky, A. M. Perelomov, 
    Phys. Rep. \textbf{94}, 6 (1983).
    \\ \textit{Quantum Integrable Systems Related to Lie Algebras.}
 \\
	A. P. Polychronakos, Les Houches Lectures, 1998,
	hep-th/9902157.
    \\ \textit{Generalized statistics in one dimension.}

\bibitem{1959-Schechter}
    S. Schechter, \textit{On the inversion of certain matrices},
    Mathematical Tables and Other Aids to Computation, 1959; Vol. \textbf{13}, no. 66., pp. 73-77.

\bibitem{Sutherland-book}
	B. Sutherland, \textit{Beautiful Models: 
	70 Years Of Exactly Solved Quantum Many-Body Problems}, 
	World Scientific, (2004).

 

\end{thebibliography}
\end{document}